\documentclass[letterpaper]{article}

\usepackage{natbib,alifeconf}  %% The order is important

\usepackage{amsmath}
\usepackage{amssymb}
\usepackage{hyperref}
\usepackage{xcolor}
\definecolor{revise}{rgb}{0, 0, 0}
\usepackage{algorithm,algpseudocode}

\title{ALIFE2022 template}

\title{Dynamics of niche construction\\in
adaptable populations evolving in diverse environments}
\author{Eleni Nisioti$^{1}$ and Clément Moulin-Frier$^{1}$ \\
$^1$~Flowers Team, Inria and Ensta ParisTech, Bordeaux, France \\
eleni.nisioti@inria.fr} % email of corresponding author

\begin{document}
\maketitle

\begin{abstract}
% Abstract length should not exceed 250 words
In both natural and artificial studies, evolution is often seen as synonymous to natural selection.
Individuals evolve under pressures set by environments that are either reset or do not carry over significant changes from previous generations.
Thus, niche construction (NC), the reciprocal process to natural selection where individuals incur inheritable changes to their environment, is ignored. 
Arguably due to this lack of study, the dynamics of NC are today little understood, especially in real-world settings.
In this work, we study NC in simulation environments that consist of multiple, diverse niches and populations that evolve their plasticity, evolvability and niche-constructing behaviors.
Our empirical analysis reveals many interesting dynamics, with populations experiencing mass extinctions, arms races and oscillations~\footnote{We provide an online repo for reproducing our simulations at \url{https://github.com/eleninisioti/NicheConstructionModel/tree/main}}.
To understand these behaviors, we analyze the interaction between NC and adaptability and the effect of NC on the population's genomic diversity and dispersal, observing that NC diversifies niches.
Our study suggests that complexifying the simulation environments studying NC, by considering multiple and diverse niches, is necessary for understanding its dynamics and can lend testable hypotheses to future studies of both natural and artificial systems.
\end{abstract}

\section{Introduction}
Biological organisms and their environments share a reciprocal relationship: organisms survive and reproduce under selection pressures present in their habitats and environments are modified by their inhabitants, with changes being inherited to the next generation and accumulating with evolutionary time~\citep{odling-smee2003NicheConstructionNeglected}.
The first process, natural selection, became the cornerstone of evolutionary theory in the early 20th century~\citep{provine2001OriginsTheoreticalPopulation} and, in both natural and artificial life studies, is often seen as synonymous to evolution.
The second, niche construction (NC), was characterized as the neglected process in evolution in the early 2000s~\citep{odling-smee2003NicheConstructionNeglected}, as evolutionary theory assumed that NC's effect on selection pressures is negligible.
Since then, studies of natural populations have shown 
that NC often affects selection pressures by helping species protect themselves from environmental uncertainty and accelerates evolution by complexifying environments~\citep{odling-smee2003NicheConstructionNeglected,schwilk2003FlammabilityNicheConstruction,boivin2016EcologicalConsequencesHuman}.
By now, NC is, at best, the elephant in the room of evolutionary synthesis: despite evidence for its existence, our understanding of it is too limited to enable its study in settings capturing the complexity of the real world.

Recent hypotheses studying major events in evolution point to environmental complexity~\citep{maslinSynthesisTheoriesConcepts2015,pottsHomininEvolutionSettings2013,derricoIdentifyingEarlyModern2017}.
For example, the birth of our own lineage in East Africa occurred amidst large climatic instability that fragmented the landscape into patches of land differing in resource availability and separated by large lakes~\citep{MaslinEastAfricanClimate2014}.
Other hypotheses further emphasize the ability of natural landmarks such as deserts and oceans to form barriers that isolate populations~\citep{larrasoanaDynamicsGreenSahara2013}.
A similar story is unfolding in artificial life: under Quality-Diversity optimization~\citep{pugh2016QualityDiversityNew}, the search space is divided into behavioral niches to evolve diverse solutions, a paradigm that challenged the dominant approach of training a single agent, as it proved more robust in real-world settings~\citep{cullyRobotsThatCan2015}.

When in a heterogeneous environment, organisms can survive by: a) specializing in a niche to out-compete others, paying the cost that they are out-competed in other niches~\cite{nisiotiPlasticityEvolvabilityEnvironmental2022a,groveEvolutionDispersalClimatic2014a} b) becoming \textit{phenotypically plastic}: a plastic individual adapts its phenotype to its environment without genetic change and survive in diverse niches within its lifetime.
On the downside, plasticity comes with fitness costs, so that plastic individuals are out-competed by specialists in their preferred niche ~\citep{groveEvolutionDispersalClimatic2014a,nisiotiPlasticityEvolvabilityEnvironmental2022a} c) becoming highly evolvable: a high mutation rate enables quick adaptation at an evolutionary scale but comes at the cost of increased deleterious mutations~\citep{lynchGeneticDriftSelection2016,nisiotiPlasticityEvolvabilityEnvironmental2022a}. As both plasticity and evolvability enable adaptation we jointly refer to them as adaptability.

Although limited, our understanding of NC shows that it is influenced both by the adaptability of populations and the diversity of environments.
For example, processes such as the invention of agriculture~\citep{zohari1986OriginEarlySpread} and cultural innovation ~\citep{mesoudiWhatCumulativeCultural}
are both influenced by geography and our impressive social learning abilities~\citep{diamond98,miglianoHuntergathererMultilevelSociality2020,miglianoOriginsHumanCumulative2022}.
Explanations for this rely on the inter-play between a population's connectivity and its ability to accumulate solutions: spatial heterogeneity enforces isolation and, therefore, diversity within a population, while adaptability ensures that environmental barriers can be crossed and solutions spread, causing an intensification of NC~\citep{cantorInterplaySocialNetworks2013,derexPartialConnectivityIncreases2016,nisioti2022SocialNetworkStructure}.

In this work, we study the evolution of NC in environments divided in diverse niches
where populations evolve their adaptability and niche-constructing behavior.
Each niche is characterized by its environmental state, which determines how many agents it can fit, and consists of two components: the intrinsic state (which can for example model climatic variations uncontrollable by the agents) and the niche-constructed state, which is the product of niche-constructing individuals inhabiting the niche.
The niche-constructed state is also inherited from the previous generation with some decaying applied to capture the fact that ecologically-inherited artifacts cannot persist indefinitely.
{\color{revise} Modeling NC as a process that changes the capacity of the environment is common~\citep{laland1999EvolutionaryConsequencesNiche,krakauerDiversityDilemmasMonopolies2009} and finds inspiration in natural behaviors, such as nest-constructing species.}
Genomes consist of four genes: preferred environmental state (the state at which their fitness is highest), plasticity (the variance in environmental state they can tolerate without a large impact on fitness), evolvability (the mutation rate) and niche construction (the amount by which they change the environmental state of a niche they inhabit).
We consider two different mechanisms for selecting which agents will reproduce: a) under \textit{global competition} we select them with a probability proportional to their average fitness across the niches they can survive in and reproduction stops
when the environment's capacity is reached b) under \textit{local competition} we consider reproduction in each niche independently: an agent is selected based on its fitness in each niche and reproduction stops when the capacity of that niche is filled.
Thus, agents benefit from surviving in multiple niches.
{\color{revise}
Our models of local and global competition have their equivalents both in previous studies in evolutionary computation (with Quality-Diversity algorithms representing local competition and classical algorithms global) and in the study of natural populations (where local adaptation considers that populations of the same species inhabiting different niches experience different natural selection pressures~\citep{leimu2008MetaAnalysisLocalAdaptation}).}

An agent may benefit from NC by: a) increasing the capacity of a niche, {\color{revise}for example through nest-building}, thus, reducing competition for resources b) reducing the capacity of a niche thus making it less desirable to others. {\color{revise}
This behavior, which we refer to as negative niche construction, is exemplified in nature by pines, which spread their needles to increase the probability of fire and out-compete not fire-resistant species~\citep{schwilk2003FlammabilityNicheConstruction}} c) bringinf the environmental state closer to its preferred niche and further away from others {\color{revise} For example, aquatic earthworms can inhabit earth soil only by changing its consistency~\citep{ExtendedOrganismPhysiology}}. But NC also comes at a cost: a) increasing the capacity of a niche may invite other species in and increase competition b) decreasing its capacity may make it uninhabitable c) changing the environmental state will require adaptability, which comes at its own cost d) multiple agents co-inhabiting a niche can create high environmental uncertainty that can lead to mass extinctions if adaptability cannot increase enough to respond to it{\color{revise}, a case that feels most relevant to our own evolution~\citep{boivin2016EcologicalConsequencesHuman,ScaleUniversalLaws}}.

Taking into account these complex interactions between environmental heterogeneity, population adaptability and niche construction, we believe that studies of NC should depart from simplified settings with a single niche~\citep{laland1999EvolutionaryConsequencesNiche,suzukiEffectsTemporalLocality,chibaEvolutionComplexNicheConstructing2020} and non-plastic populations~\citep{laland1999EvolutionaryConsequencesNiche,krakauerDiversityDilemmasMonopolies2009}. As we show in this work, rich models can offer interesting hypotheses about the dynamics of NC, such as that:
\begin{itemize}
    \item having multiple niches is necessary for avoiding mass extinctions. We observed that, in a single niche, environmental uncertainty induced by NC becomes too high for adaptability to cope with;
    \item NC promotes adaptability and populations adapt differently depending on the selection mechanism: under global competition NC leads to higher plasticity while under local to higher evolvability;
    \item the population may niche-construct negatively for prolonged periods of time, making niches uninhabitable. Agents deal with this by evacuating the niche until it becomes inhabitable due to the decay of NC;
    \item under local competition NC increases genomic  and environmental diversity
\end{itemize}

\section{Related works}
Previous studies have primarily modeled NC in two different ways: as an increase in the amount of resources that increases the environment's capacity~\citep{laland1999EvolutionaryConsequencesNiche,krakauerDiversityDilemmasMonopolies2009} and as a direct increase or decrease in fitness that does not affect capacity~\citep{suzukiEffectsTemporalLocality}. We follow the first approach but also allow reductions in capacity. NC studies were at first primarily theoretical, employing differential equations to predict how NC affects evolution~\citep{laland1999EvolutionaryConsequencesNiche,krakauerDiversityDilemmasMonopolies2009}. Such models lent intuitive insights, such that NC can emerge without direct selection~\citep{laland1999EvolutionaryConsequencesNiche} and that the ability to monopolize niches enables the emergence of NC~\citep{krakauerDiversityDilemmasMonopolies2009} but come with limitations, such as the fact that niches are identical and agents are not plastic.

\citet{suzukiEffectsTemporalLocality} study the co-evolution of plasticity and NC in an agent-based model and show that populations alternate between phases of high plasticity and NC, provided that agents niche-construct sequentially.
This model differs from ours as the environment consists of a single niche, NC does not modify its capacity and evolvability is constant.
Interestingly, when \citet{suzukiEffectsTemporalLocality} applied NC in parallel these patterns disappeared due to agents canceling out each other's behavior. 
Instead, the patterns in our analysis occur with NC applied in parallel.
As we show, this becomes possible due to the presence of multiple niches that stabilize NC.  
\citet{nisiotiPlasticityEvolvabilityEnvironmental2022a} studie the co-evolution of plasticity and evolvability in an environment with multiple niches, but do not consider NC. 

Recent advents in deep reinforcement learning (DRL) and neuroevolution have enabled the study of complex behaviors in simulations, such as foraging and tool-use~\citep{NIPS2017_2b0f658c,gautier2023EcoevolutionaryDynamicsNonepisodic,bakerEmergentToolUse2020}. \citet{chibaEvolutionComplexNicheConstructing2020} leverage such techniques to study NC in a 2D environment where agents can construct artifacts useful for avoiding predation. Our study can lend insights to this direction, as we can view DRL as the mechanism that underlies the behavioral plasticity considered in our model.

\section{Modeling and methodology}

We now separately discuss our model of the environment and genomes, the evolutionary algorithm and the set of metrics we use to monitor evolution.

\begin{figure}
\begin{minipage}{\columnwidth}
    \includegraphics[width=0.8\columnwidth]{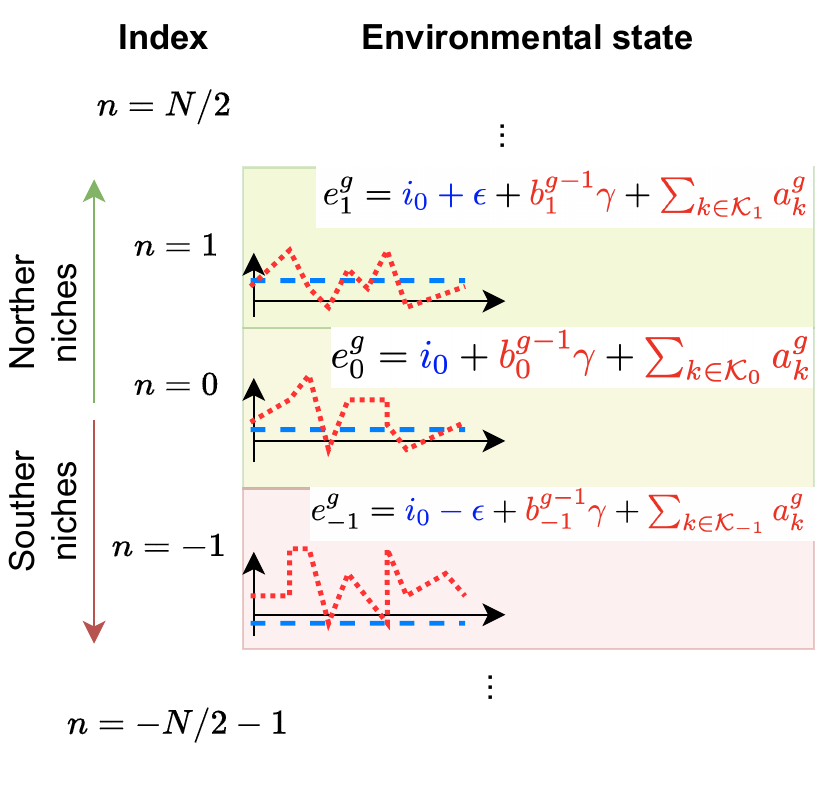}
\end{minipage} 
\caption{Our model of the environment: each niche $n$ is characterized by its environmental state $e_n^g$ which is the sum of the intrinsic state $i_n$ (in blue) and the niche-constructed state $b_n^g$ (in red). $i_n$ depends on the state of the reference niche $(e_0)$, the offset $\epsilon$ and $n$. $b_n^g$ is the sum of the niche-constructed state at the previous generation discounted by $\gamma$ plus the sum of the niche-constructing behavior of agents that reproduced in this niche in the current generation.}
 \label{fig:env}
\end{figure}

\subsection{Modeling the environment}\label{sec:env_model}
The environment, illustrated in Figure \ref{fig:env}, is divided into $N$ niches arranged in a simple latitudinal model: we consider a reference niche at $n=0$, $N/2$ ``northern" niches indicated with positive indexes $n \in (0,N/2]$  and $N/2-1$ ``southern" niches with negative indexes $n \in (-N/2,0)$ . Each niche is characterized by its environmental state $e_n^g$ which is the sum of two elements:
\begin{itemize}
    \item an intrinsic state $i_n$ that remains constant with evolutionary time and depends on the location of the niche. Specifically, the state of niche $n$ at generation $g$ is $i_n = i_0 + \epsilon \cdot n$, where $i_0$ is the state of the reference niche and $\epsilon$ is a constant capturing the difference between adjacent niches. 
    \item the niche-constructed state $b_n^g$, capturing the modifications that agents inhabiting the niche cause. These modifications are carried over generations but are discounted by a factor $\gamma$.
    Formally, the niche-constructed state $b_n^g$ of niche $n$ is equal to $b_n^{g-1}\cdot\gamma$, with $\gamma<1$, plus the total amount of NC applied by all agents that reproduced in this niche in generation $g$. We denote this latter amount as $\sum_{k \in \mathcal{K}_n} a_k^g$, where $\mathcal{K}_n$ is the subset of all agents that reproduced in niche $n$ and $a_k^g$ is the amount of niche construction by a single agent (we later explain how agents niche-construct).
\end{itemize}
Thus, the general equation describing the evolution of niche $n$ with generations $g$ is:
\begin{align}
    b_n^g &=  b_n^{g-1} \cdot \gamma + \sum_{k \in \mathcal{K}_n} a_k^g \nonumber \\
    e_n^g &= i_0 + \epsilon \cdot n + b_n^g
\end{align}

As we explain later, the environmental state determines the fitness of an agent based on its genome and  the capacity of the niche $c_n^g$ as: $c_n^g=e_n^g C_N$, where $C_N$ is the reference niche capacity.
Thus, higher environmental states can support larger populations and are termed ``high-quality".
To ensure that the maximum population size is independent of the number of niches we define $C_N=C_{\text{ref}}/N$, where $C_{\text{ref}}$ is equal to the desirable maximum population size.
An assumption of this model is that there is spatial smoothness, i.e, nearby niches are similar.
Note that $C_{\text{ref}}$ is independent of niche-constructing behavior: by niche-constructing the population can exceed this capacity. We, therefore, further bound the population size by randomly discarding agents if the population exceeds a value $K_{\text{max}}$.

\subsection{Modeling the genome}\label{sec:genome_model}
\begin{figure}
\begin{minipage}{\columnwidth}
\centering
    \includegraphics[width=0.9\columnwidth]{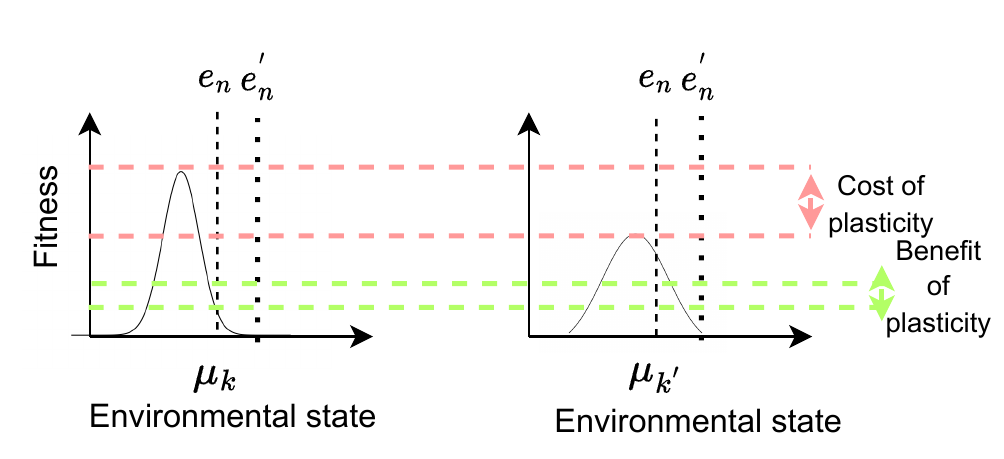}
\end{minipage} 
\caption{Modeling plasticity as a normal distribution $\mathcal{N}(\mu_k,\sigma_k)$. A non-plastic individual ($k$) has small $\sigma_k$ and a high peak at their preferred niche, while a plastic individual ($k^{'}$) has large $\sigma_k$ and a lower peak at their preferred niche. Fitness in a given niche $n$ is computed as the probability density function of the distribution at the environmental state $e_n$. This figure also illustrates the cost and benefit of plasticity, assuming that $\mu_k=\mu_k^{'}$. If $e_n=\mu_k$ (the actual environmental state is identical to the preferred niche of both individuals) the plastic individual has lower fitness (cost of plasticity). If $e_n$ differs significantly from  $\mu_k$ (the actual environmental state differs from the preferred one) the plastic individual has higher fitness (benefit of plasticity).}
 \label{fig:plasticity}
\end{figure}

We model plasticity through tolerance curves, a tool developed in ecology~\citep{doi:10.1086/284635} and previously employed in simulation environments~\citep{groveEvolutionDispersalClimatic2014a,nisiotiPlasticityEvolvabilityEnvironmental2022a}.
A tolerance curve is a normal distribution with mean $\mu_k^g$, indicating the environmental state of highest fitness for an individual $k$ at generation $g$, called the preferred state, and standard deviation $\sigma_k^g$ that captures how quickly the fitness of the genome drops as the environmental state varies from its preferred state.
Genomes with large $\sigma_k^g$ are indicative of plastic individuals (we illustrate the tolerance curves of a plastic and a non-plastic individual in Figure \ref{fig:plasticity}).  

To model niche construction we introduce a niche-constructing gene $a_k \in [-N/K_{\text{max}}, N/K_{\text{max}}]$ where we bound the amount an agent can niche-construct to ensure that agents inhabiting large environments have limited capabilities. 
The niche-constructing gene is expressed in the niche an agent reproduces in by modifying its environmental state.
Thus, at each generation, niche $n$ is constructed by an amount $\sum_{k \in \mathcal{K}_n} a_k$, where  $\mathcal{K}_n$ denotes the subset of agents that reproduced in it. 
The genome $o_k^g$ also includes the mutation rate $r_k^g$. Thus the complete form of a genome is $o_k^g=[\mu_k^g, \sigma_k^g, r_k^g, a_k^g]$ and, upon reproduction, it mutates as:

\begin{align} 
\mu_k^{g+1} = \mu_k^{g} + \mathcal{N}(0, r_k^{g}) \nonumber \\
\sigma_k^{g+1} =\sigma_k^{g} + \mathcal{N}(0, r_k^{g}) \nonumber \\
a_k^{g+1} = [c_k^{g} + \mathcal{N}(0, r_a)]_{-N/K_{\text{max}}}^{ N/K_{\text{max}}} \nonumber \\
r_k^{g+1} = r_k^{g} +\mathcal{N}(0, r_k^{g}) \label{eq:mutation}
\end{align}
where $\mathcal{N}(x,y)$ denotes a normal distribution with mean $x$ and variance $y$ and we have highlighted that the niche-constructing gene is bounded. Note also that, for stability reasons, we employ a different mutation rate for the niche-constructing gene, $r_a$, that remains constant.

\subsection{Global and local selection}

At the end of a generation agents are selected for crossover based on their fitness. Each chosen individual has two offspring and the next generation consists only of offspring. To compute the fitness of an individual $k$ in generation $g$ we first detect the niches in which it can survive as: 
\begin{align}\label{eq:survive_niche}
{n \in \{1, \cdots, N\} \quad | \quad e_n^g \in [\mu_{k}^g - 2\sigma_{k}^g,\mu_k^g + 2\sigma_k^g]}
\end{align} and
compute its fitness in each one of them as $f_{k,n}^g=pdf(\mu_k^g, \sigma_k^g, e_n^g)$, where $pdf$ denotes the value of the normal probability density function with mean $\mu_k^g$, and  variance $\sigma_k^g$ at location $e_n^g$. We study two selection mechanisms:
\begin{itemize}
    \item under global selection all agents in the population are ranked based on their {\color{revise} average fitness across the niches they can survive in} and reproduce with a probability proportional to it. {\color{revise} Agents reproduce until the capacity of the environment is filled. }
    \item under local selection we apply the same criterion but independently for each niche: we detect which agents survive in a niche and reproduce them with a probability proportional to their fitness. Thus, agents that can survive in multiple niches have higher chances of reproduction under this mechanism. Again, agents reproduce in a niche until its capacity is filled. 
\end{itemize}

Thus, at the end of a generation a new population is formed that consists of the offspring of agents from the previous population that were chosen for reproduction, based on their fitness and the environment's capacity. 
Note that the niche is not inherited from parent to offspring: the offspring will inhabit the niches its genome describes (as in Eq. \ref{eq:survive_niche}). 
We present the pseudocode of our algorithm in Algorithm 1 and 2, provided in our \href{https://github.com/eleninisioti/NicheConstructionModel/tree/main}{online repo} due to limited space.

\subsection{Metrics}

In addition to population-wide averages of the genome values and environment-wide averages of the intrinsic and niche-constructed states we monitor the following metrics:

\begin{enumerate}
    \item $X^g=\sum_k X_k^g$, the number of extinctions. We denote the survival of individual $k$ in niche $n$ at generation $g$ as a binary variable:
    \begin{align}\label{eq:survival}
        s_{k,n}^g = (e_{n,g} \in [\mu_{k}^g - 2\sigma_{k}^g,\mu_k^g + 2\sigma_k^g])
    \end{align}
    Thus, an individual goes extinct ($X_k^g=1$) if $\sum_n^N s_{k,n}^g$ is zero and survives ($X_k^g=0$) if $\sum_n^N s_{k,n}^g$ is positive. 
    \item $V^g_{\mu}$, the diversity of the population defined as the standard deviation of the population's preferred state, namely
    \begin{align}\label{eq:diversity}
        V^g_{\mu} = \sigma_{\mu^g}
    \end{align} 
    This metric captures the genetic diversity of the population computed for the gene of preferred state.
    \item $D^g$, the dispersal of the population, computed as the number of niches over which at least one individual survives for a temporal window of at least $w$ generations. Formally, $D^g=\sum_{n=1}^N d_{n,w}^g$,  where $d_{n,w}^g$ denotes the persistence of the population in a given niche for the required time window and is computed as
    \begin{align}
        d_{n,w}^g = \begin{cases} 1 \quad \text{if} \sum_{g^{\prime}=g-w}^g s_{n}^{g^{\prime}}=w \\
        \text{0} \quad \text{otherwise}
        \end{cases}
    \end{align}
    where $s_{n}^g$ is indicates the survival of at least one individual in a given niche and is computed as 
    \begin{align}
        s_{n}^g = \begin{cases}
        1 \quad {if} \sum_{k}^K s_{k,n}^g > 1 \\
        0 \quad \text{otherwise}
        \end{cases}
    \end{align} with $s_{k,n}^g$ defined in Eq. (\ref{eq:survival}).
    \item $H^g$, the competition for reproduction within the population. We count, for each niche, the number of agents that survived in it and were fit enough to be chosen for reproduction but did not reproduce because its capacity was reached and, then, sum over all niches. 
\end{enumerate}

\section{Results}
We now examine the behavior of agents under different settings.
First, we compare behaviors under global competition between environments with different number of niches.
After, we consider only heterogeneous environments ($N=100$) niches and we contrast the behavior across two dimensions: a) whether NC takes place or not (where we force all niche-constructing genes to be zero) and b) whether selection is local or global.
In all simulations we set the discount factor $\gamma$ to 0.5, the intrinsic state of the reference niche $i_0$ to 0.6, $\epsilon$ (the difference between adjacent niches) to 0.01, the reference capacity to $C_{\text{ref}}=1000$, the maximum population size to $K_{\text{max}}=5000$ and the mutation rate of the niche-constructing gene, $r_c$, to 0.0003.
{\color{revise} As we discuss later it would be interesting to study the effect of some of these hyper-parameters.
For this study we note that the values of $C_{\text{ref}}, K_{\text{max}}$ where chosen to limit computational complexity, increasing them further should not qualitatively change our conclusions. Also regarding $r_c$, setting it to too low a value will disable NC and setting it too high may destabilize NC.} 
We performed ten independent trials and present in plots median values and $95\%$ confidence intervals.
When studying niche-constructing behaviors we do not average across trials but present them individually, as differences would average out and conceal information.

\subsection{Spatial heterogeneity stabilizes niche construction}

\begin{figure}
    \centering \includegraphics[width=0.7\columnwidth]{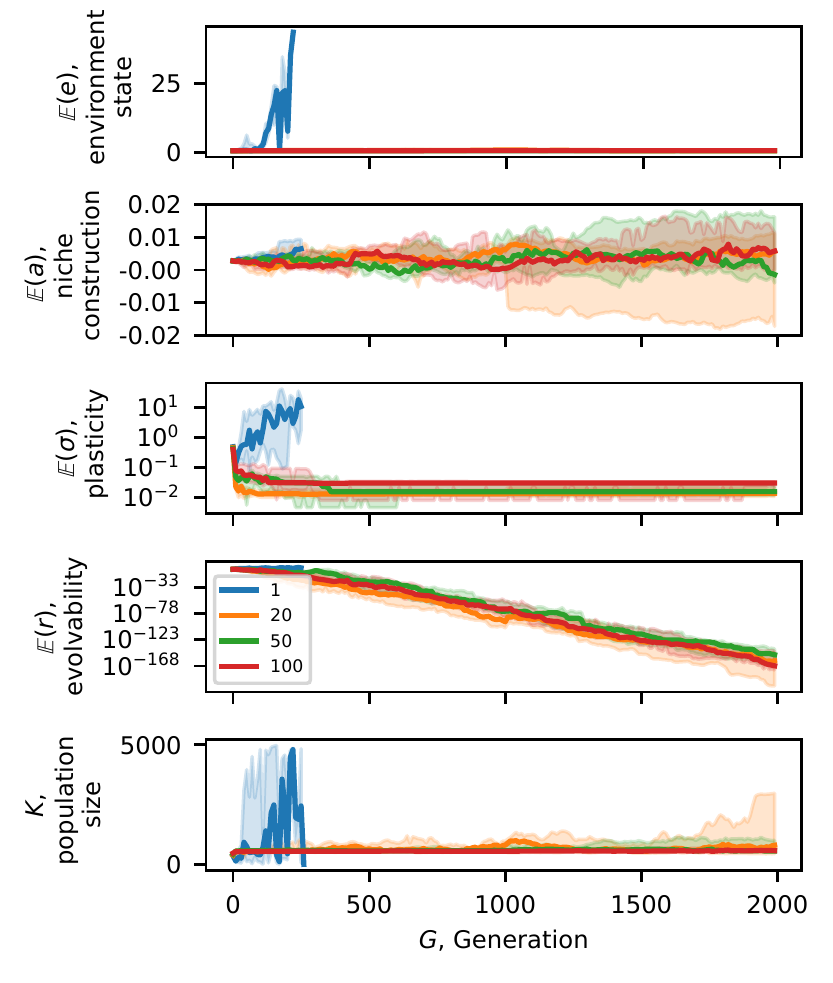}
    \caption{Comparison between a homogeneous ($N=1$) and a heterogeneous environment ($N=100$) under global competition}
    \label{fig:one_niche}
\end{figure}

In Figure \ref{fig:one_niche} we compare the behavior under global competition for
different number of niches ($N \in \{1,50,100\}$ for a single trial.
We observe that, for $N=1$, the agents are niche-constructing positively (third row), which pushes the environment state to higher values and more variability (first row). We see that the the population reacts by adapting its preferred niche (second row), increasing its plasticity (fourth row) and keeping its evolvability high (fifth row). Despite that, the population goes extinct around generation 250, after experiencing many booms and busts (sixth row). This behavior was consistent across all trials for $N=1$, with half of the trials experiencing a mass extinction due to negative NC and half due to positive NC (as in Figure \ref{fig:one_niche}).

In the heterogeneous environments, on the other hand, the population survives for the whole simulation. We see that the average NC among agents ($\mathbb{E}(a)$) is similar to that of the homogeneous environment but that this leads to less niche construction in the environment (evident through the environmental state). This is because the population in the heterogeneous environment spreads over multiple niches and often migrates collectively out of a niche (we will take a look at this behavior in the next section), leading to a decrease to the accumulation of NC due to its discounting. In contrast, when $N=1$ NC accumulates in a single niche. We also observe that NC in $N=100$ does not increase continuously but experiences oscillations, which suggests that some agents are niche-constructing negatively or reducing their positive NC. As a result, the environment experiences lower oscillations which enables the population to reduce its evolvability. Plasticity remains relatively high but much lower than in the homogeneous environment.
{\color{revise} We also measured the number of mass extinctions for environments with different number of niches and observed that even a small number of niches is useful and increasing the number of niches continuously decreases the probability of extinction (10/10 went extinct for $N=1$, 4/10 for 20, 3/10 for 50 and 2/10 for 100).}
Thus, the presence of multiple niches was necessary for the survival of the niche-constructing population.
Next, we will look into how adaptability interacts with NC to give rise to this behavior. As we hypothesize that the presence of niches matters, we will focus on how agents disperse and niche-construct differently in different niches.

\subsection{Niche construction promotes adaptability}
 \begin{figure}[t]
    \centering
\includegraphics[width=0.7\columnwidth]{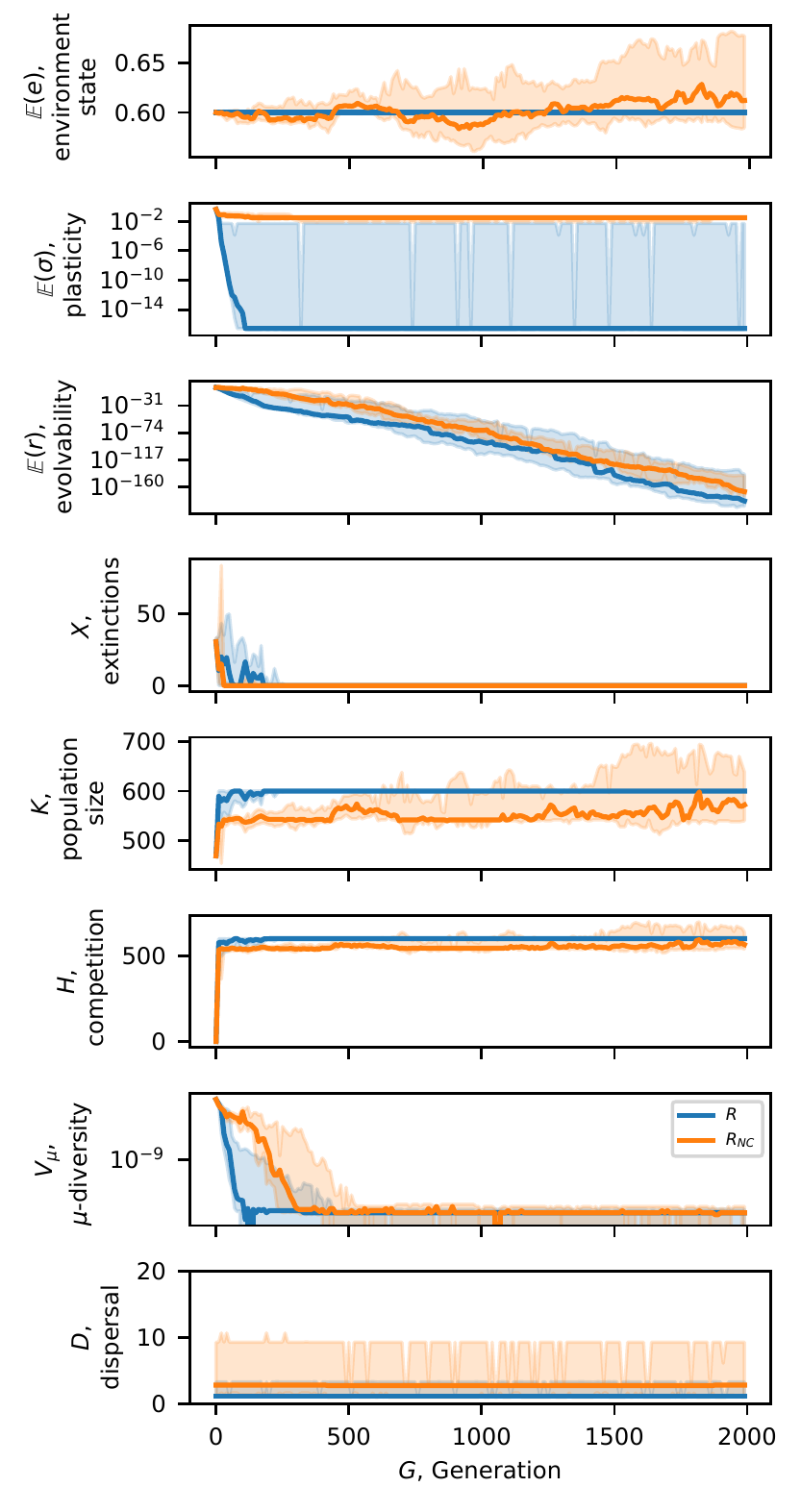}
    \caption{Comparison between populations with ($R_{NC}$) and without NC ($R$) under global competition.}
    \label{fig:F_average}
\end{figure}

\begin{figure}[t]
    \centering
\includegraphics[width=0.7\columnwidth]{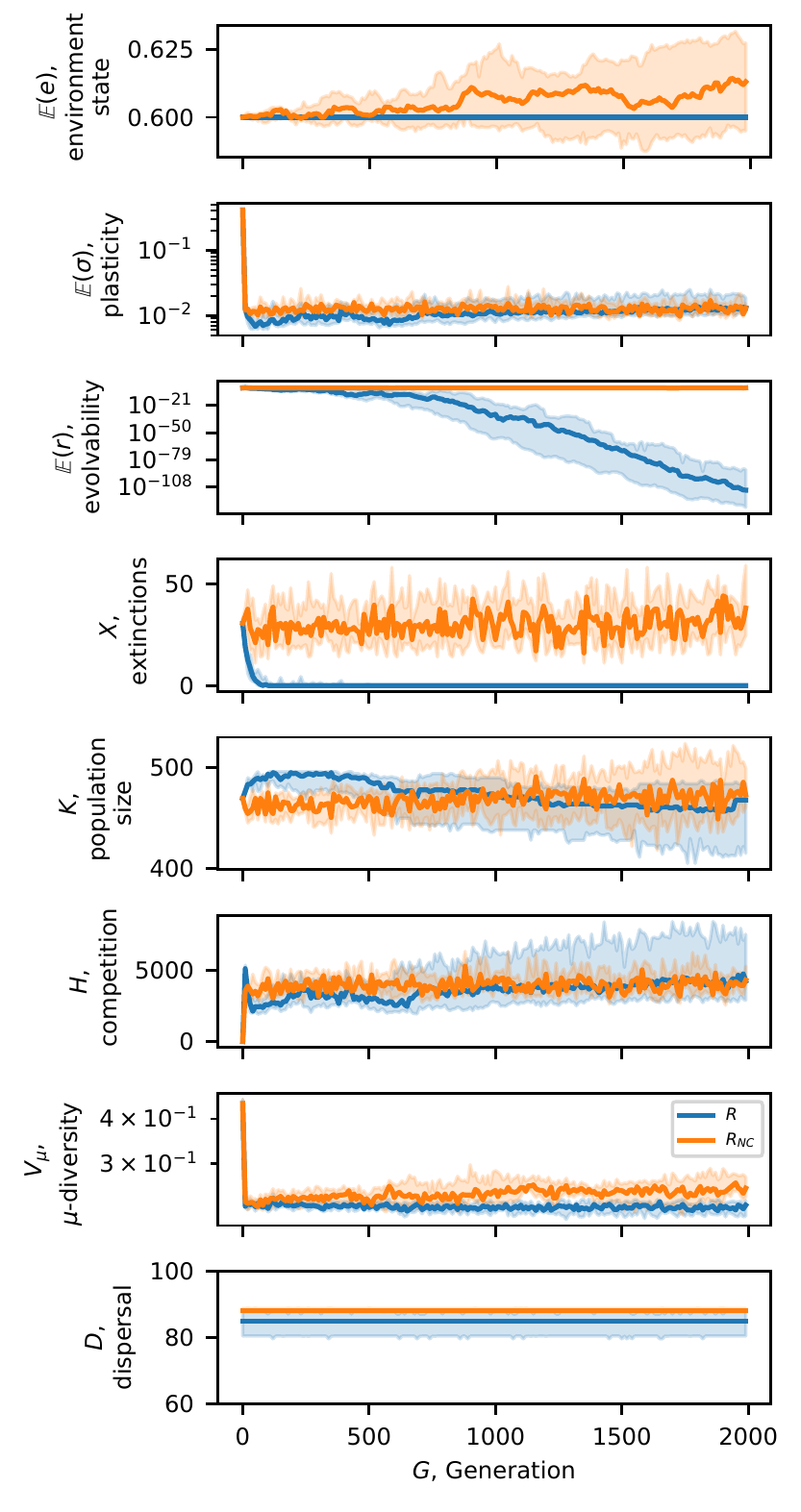}
    \caption{Comparison between populations with ($R_{NC}$) and without NC ($R$) under local competition.}
    \label{fig:NF_average}
\end{figure}

We now compare a niche-constructing population (denoted as $R_{NC}$) and a population that cannot niche -construct ($R$).
We examine populations under global selection in Figure \ref{fig:F_average} and under local selection in Figure \ref{fig:NF_average}.
We observe that both populations increase their adaptability but do so differently. 

Under global selection, the $R_{NC}$ population increases its plasticity but keeps its evolvability low.
Dispersal is a bit higher and genomic diversity is initially high but, by the middle of the simulation, reaches the low level of the $R$ population. We also see that $R_{NC}$ has a smaller population size early in the simulation.
As extinctions are low, this is due to NC reducing the capacity of the environment. Near the end of the simulation we see that $R_{NC}$ reaches and, in some trials, surpasses the size of the $R$ population. This indicates that positive NC has increased the environment's capacity.

Under local selection $R_{NC}$ has very high evolvability but plasticity is at similar level to $R$.
We observe that some extinctions persist throughout evolution but the population size is relatively stable. 
{\color{revise} Why did the niche-constructing population prefer to adapt through evolvability rather than plasticity?
% With plasticity at 0.01 both $R_{NC}$ and $R$ can inhabit at most two niches.
% NC increases uncertainty in the environment, so either plasticity or evolvability must be increased. 
As we will see more closely in our analysis of specific trials, agents in a niche-constructing population manage to coordinate their NC behavior within a niche.
This would not have been possible under high plasticity, as agents would stochastically niche-construct in different niches and increase environmental uncertainty.
}

% An interesting observation is that, under NC, competition is lower, which shows that it is less likely for niches to overflow and discard fit agents. This could be explained by the fact that genomic diversity and dispersal are higher: agents are diverse enough to disperse and not interfere with each other and can use NC within their niche to ensure that capacity is not reached.
% % Thus, results suggest that introducing NC results in increased adaptability. 
% To understand how this comes about we will now look into specific trials and analyze the diversity of collective behaviors that we observed under different conditions.

\subsection{Niche construction can cause an arms race}

\begin{figure}[t]
    \centering
    % \begin{flushleft}\includegraphics[width=0.8\columnwidth]{images/F_trial1.pdf}
    % \end{flushleft}   \includegraphics[width=0.8\columnwidth]{images/heatmap_F_trial1}
    \includegraphics[width=0.75\columnwidth]{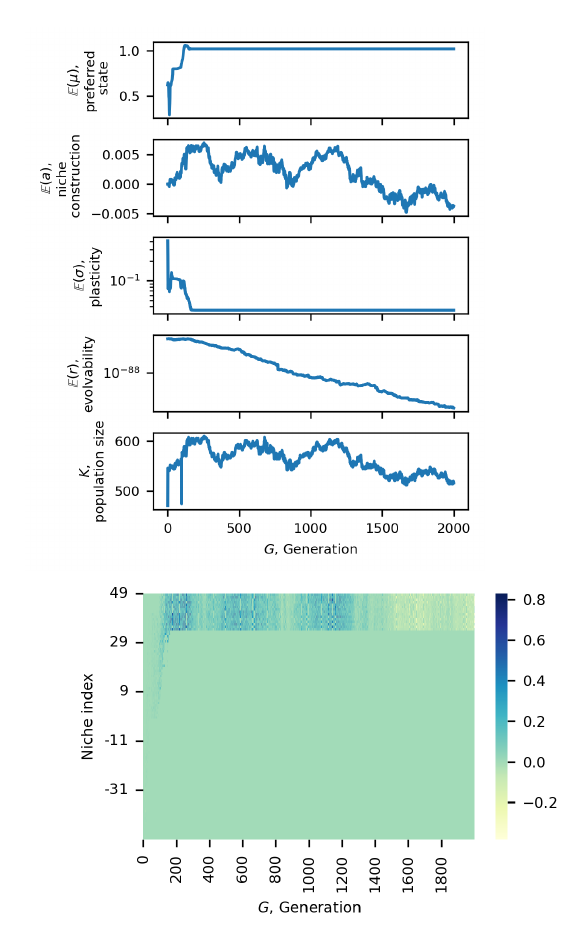}
    \caption{Analyzing one trial for global competition with niche-constructing population where an arms race emerged (Top) evolution of metrics (Bottom) Heatmap with rows corresponding to niches, columns to generations and value/color of cell to the sum of niche construction of all agents in a given niche and generation.}
    \label{fig:F_trial1}
\end{figure}

In Figure \ref{fig:F_trial1} we analyze one of the trials for the niche-constructing population under global competition through heatmaps that show the amount of NC at each generation and niche.
Populations are initialized with diverse genomes, so that at the beginning they are dispersed in all niches.
Then, global competition leads to a decrease in plasticity and quickly wipes out the diversity of the population, gathering all agents in the same narrow stripe of niches.
Around generation 50, the population starts positively niche-constructing in the middle niche, then starts moving to the north and, once it reaches the end of the environment (around generation 150), stays there and oscillates between phases of positive and negative NC. On the top plot, we see that this population has adapted its preferred state to the last niche and that as soon it reaches it, plasticity stops decreasing and the population size stops increasing.
The trial depicted in Figure \ref{fig:F_trial1} is representative of 4 out of the 10 trials.
In another five trials the same pattern occured, but in the south instead of the north.
In contrast, we observed that populations without NC do not gather at the edge of the environment but stay in a random niche near the center.
Thus, NC under global competition leads to a "geographic" arms race due to the following dynamics: once the agents have been gathered to a narrow stripe of niches, they begin niche-constructing, initially randomly. Then, NC will, by random chance become slightly positive or negative. As all agents are competing in the same niche, this will force them to adapt their preferred state towards the direction that the sum of agents is constructing to. This enables inhabiting the adjacent niche, where the agents will niche-construct towards the same direction. Eventually, the population will reach the end of the environment. Why doesn't the population continue increasing its niche construction to the same direction instead of experiencing cycles? We believe that this is not possible as the mutation rate is low and, therefore, the preferred niche cannot be adapted anymore.

\subsection{Negative niche construction reduces competition}
\begin{figure}[t]
    \centering
    \includegraphics[width=0.75\columnwidth]{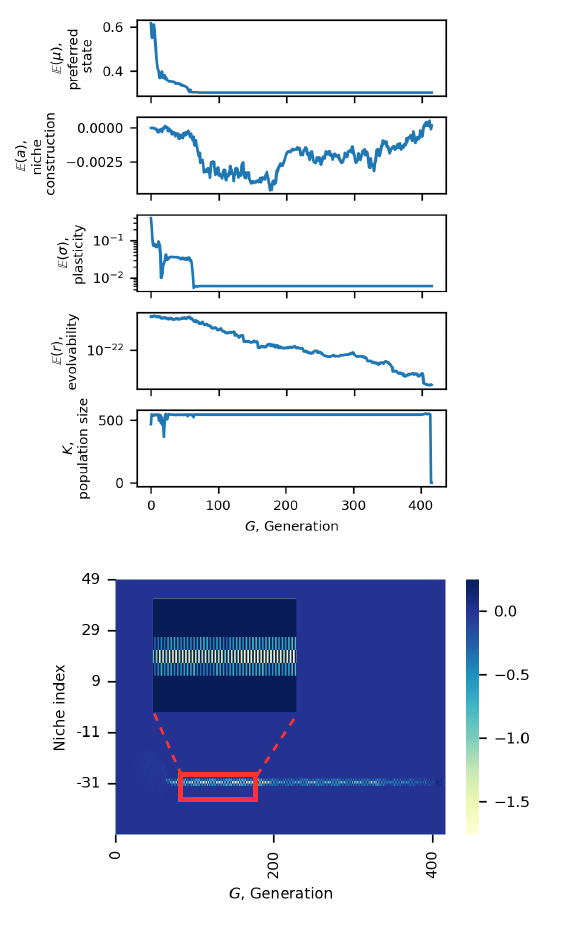}
    \caption{Analyzing one trial for global competition with niche-constructing population where negative niche construction emerged.}
    \label{fig:F_trial4}
\end{figure}
In Figure \ref{fig:F_trial4} we analyze another trial for the niche-constructing population under global competition, where a different behavior emerged: the population niche-constructs negatively, which pushes it to the low-index niches. There we see an interesting pattern: the population is gathered in a narrow stripe of niches where it niche-constructs negatively and with increasing intensity. Then the population leaves this stripe and moves to the adjacent northern and southern niches. This is possible due to the maintained plasticity and does not require an adaptation of the preferred niche. Then, the same behavior happens until the population moves back to the previous stripe and we see many cycles of moving back and forth. This switching behavior is caused by the fact that, once the population niche constructs negatively in a niche for some time, then this niche becomes uninhabitable, so that the population needs to move out until it becomes inhabitable again. Eventually, some positive NC happens, which leads to an extinction, as evolvability is too low to enable adaptation. This indicates that negative niche construction is not stable in the long-term.

\subsection{Niche construction diversifies the environment}

\begin{figure}
    \centering
\includegraphics[width=0.75\columnwidth]{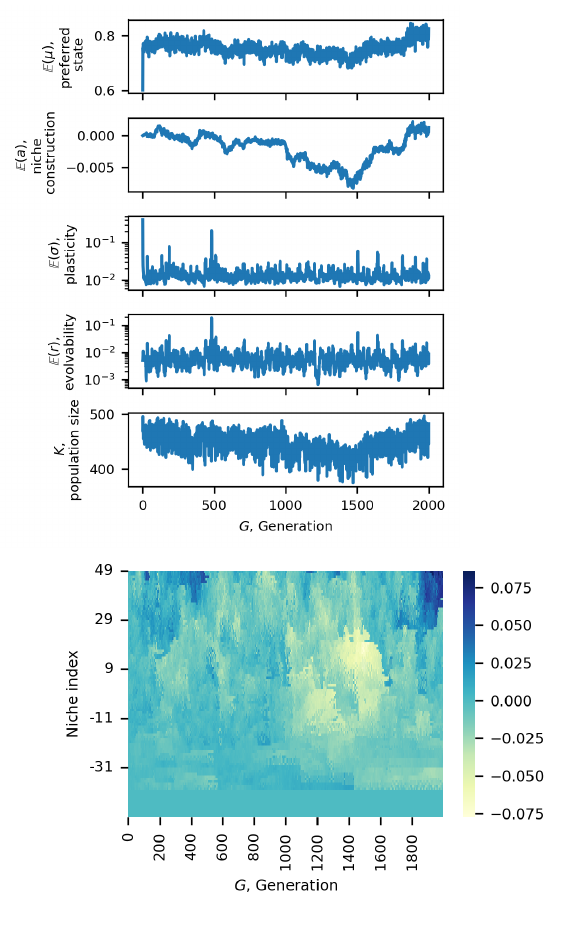}
    \caption{Analyzing a typical trial for local competition with a niche-constructing population.}
    \label{fig:NF_trial7}
\end{figure}

We now move to local selection, where we saw in Figure \ref{fig:NF_average} that NC leads to a more evolvable population.
In Figure \ref{fig:NF_trial7}, we analyze a typical trial where we see that NC is positive and high in northern niches and negative in more southern niches.
Thus, the majority lives in the north (this can be inferred from the preferred state $\mathbb{E}(\mu)=0.8$ surpassing the state of the reference niche $i_0=0.6$). The population experiences many changes, with niches often switching between being positively and negatively niche-constructed.
We also measured the variance of NC within a niche and found that it is two orders of magnitude smaller than the one under global selection.
This suggests that local selection enables agents to coordinate their niche construction within a niche, which diversifies the environment and leads to the higher genomic diversity observed in Figure \ref{fig:NF_average}.

An intriguing question is why populations under global selection adapt by increasing their evolvability but keep their plasticity at even lower levels than without NC. We believe that this is due to the interplay between NC and plasticity: very plastic agents could niche-construct in many niches, which would make it difficult to maintain low variance in NC within a niche. By monopolizing a limited number of niches and quickly changing niche through mutations these populations can better coordinate their NC.

%\subsection{Negative niche construction avoids population booms}
% are u sure it is competition? cause competition is very large in one niche. so  it's probably not competition in terms of niche construction
% then maybe competition in terms of reproduction? 
% I see that competition is lowered in the NF case, arguably due to higher diversity enabled by higher evolvability
% then, maybe it is the increased plasticity, which is enabled by multiple niches that bounds niche construction. It also enables escaping a niche
% it could also be just the fact that there is more effective climatic variability per niche because there are less agents in a niche. but I do not think this is the reason, because even a bounded number of agents can have infinite niche construction
% or it because there is negative niche construction? I think that's it. but then maybe I need to look into why negative niche construction emerges. maybe it helps with reducing competition? probably I need a controlled experiment but not sure I can have it. I could perhaps force only positive niche construction, in which case I will for sure get an explosion. Perhaps this should be an argument against works that do not consider negative NC (Krakauer? maybe the japanese?) in the sense that they do not consider nc that reduces fitness

\section{Discussion}

We have shown that the evolution of niche construction is contingent on the number of niches in the environment and the selection mechanism.
We explained this by looking into the interplay between adaptability and NC, in particular their effect on the population's and environment's diversity and dispersal patterns.
Populations are impressive in their ability to self-regulate themselves: NC remains bounded, even though we do not introduce an explicit cost for it. Populations fail to coordinate their NC and go extinct only when the environment has a single niche.

% limitations
Our empirical study could be extended in multiple ways. 
{\color{revise}First, we could study the effect of additional hyper-parameters.
For example, we could allow the intrinsic state to vary with time, following a noisy or periodic signal~\citep{nisiotiPlasticityEvolvabilityEnvironmental2022a}.
We could also consider that NC is plastic: currently an agent's gene determines its NC independently of the environmental state while we could imagine scenarios where agents niche-construct differently for different states.
Finally, we believe that our empirical observations should be tested in grounded environments with RL agents, such as grid-worlds where a population forages or avoids predation~\citep{gautier2023EcoevolutionaryDynamicsNonepisodic,chibaEvolutionComplexNicheConstructing2020}.
}

We hope that our work will contribute towards incorporating NC in future studies of collective adaptation.
We think this is particularly relevant for the Artificial Intelligence community, which recently focused on agents that can generalize to diverse environments, the importance of environmental complexity and multi-agent dynamics ~\citep{https://doi.org/10.48550/arxiv.2205.06175,jaderbergOpenEndedLearningLeadsa,nisiotiGroundingEcologicalTheory2021,moulinfrier:tel-03875448}.
As we show here, niche construction, which can be seen as a meta-learning mechanism~\citep{constantVariationalApproachNiche2018}, can prove promising in complexifying environments and population dynamics and bring us closer to artificial agents with behaviors reminiscent of natural ones.

\paragraph{Acknowledgements}
This research was partially funded by the French National Research Agency (\url{https://anr.fr/}, project ECOCURL, Grant ANR-20-CE23-0006). This work also benefited from access to the HPC resources of IDRIS under the allocation 2020-[A0091011996] made by GENCI.

%\section{Acknowledgements}

%This work was supported by NSF grant No.\ PHY-9723972.

\footnotesize
\bibliographystyle{apalike}
\bibliography{example} % replace by the name of your .bib file

\cleardoublepage

% \onecolumn

% \input{algs/evolution_short}
% \input{algs/reproduction}

\end{document}